# Strongly exchange-coupled and surface-state-modulated magnetization dynamics in Bi$_2$Se$_3$/YIG heterostructures

Y. T. Fanchiang[1], K. H. M. Chen[2], C. C. Tseng[2], C. C. Chen[2], C. K. Cheng[1], C. N. Wu[2], S. F. Lee[3*], M. Hong[1*], and J. Kwo[2*]

[1]*Department of Physics, National Taiwan University, Taipei 10617, Taiwan*

[2]*Department of Physics, National Tsing Hua University, Hsinchu 30013, Taiwan*

[3]*Institute of Physics, Academia Sinica, Taipei 11529, Taiwan*

Abstract

We report strong interfacial exchange coupling in Bi$_2$Se$_3$/yttrium iron garnet (YIG) bilayers manifested as large *in-plane* interfacial magnetic anisotropy (IMA) and enhancement of damping probed by ferromagnetic resonance (FMR). The IMA and spin mixing conductance reached a maximum when Bi$_2$Se$_3$ was around 6 quintuple-layer (QL) thick. The unconventional Bi$_2$Se$_3$ thickness dependence of the IMA and spin mixing conductance are correlated with the evolution of surface band structure of Bi$_2$Se$_3$, indicating that topological surface states play an important role in the magnetization dynamics of YIG. Temperature-dependent FMR of Bi$_2$Se$_3$/YIG revealed signatures of magnetic proximity effect of $T_c$ as high as 180 K, and an effective field parallel to the YIG magnetization direction at low temperature. Our study sheds light on the effects of topological insulators on magnetization dynamics, essential for development of TI-based spintronic devices.

Topological insulators (TIs) are emergent quantum materials hosting topologically protected surface states, with dissipationless transport prohibiting backscattering [1,2]. Strong spin-orbit coupling (SOC) along with time reversal symmetry (TRS) ensures that the electrons in surface states have their direction of motion and spin "locked" to each other [1,3,4]. When interfaced with a magnetic layer, the interfacial exchange coupling can induce magnetic order in TIs and break TRS [5-8]. The resulting gap opening of the Dirac state is necessary to realize novel phenomena such as topological magneto-electric effect [9] and quantum anomalous Hall effect [10,11]. Another approach of studying a TI/ferromagnet (FM) system focuses on spin-transfer characteristic at the interface and intends to exploit the helical spin texture of topological surface states (TSSs). Attempts have been made to estimate the spin-charge conversion (SCC) efficiency, either by using microwave-excited dynamical method [12-16] (e.g. spin pumping and spin-torque FMR) or thermally induced spin injection [17]. Very large values of SCC ratio have been reported [13,15,16]. Recently, TIs are shown to be excellent sources of spin-orbit torques for efficient magnetization switching [18].

Since the magnetic proximity effect (MPE) and spin-transfer process rely on interfacial exchange coupling of TI/FM, understanding the magnetism at the interface has attracted strong interests in recent years. Several techniques have been adopted to investigate the interfacial magnetic properties, including spin-polarized neutron reflectivity [7,19], second harmonic generation [20], electrical transport [6,21], and magneto-optical Kerr effect [6]. All these studies clearly indicate the existence of MPE resulting from exchange coupling and strong spin-orbit interaction in TIs. For device application, however, it is equally important to understand how the interfacial exchange coupling affects the magnetization dynamic of FM. For example, TIs can introduce additional magnetic damping and greatly alter the dynamical properties of FM layer, as commonly observed in FM/heavy metals systems [22-24]. The enhanced damping is visualized as larger linewidth of FMR spectra [22-24]. In TI/FM, the presence of TSS and, possibly, MPE complicate the system under study, and is still an open question to answer.

In this work we systematically investigated FMR characteristic of the ferrimagnetic insulator YIG under the influence of the prototypical three-dimensional TI Bi$_2$Se$_3$ [25]. We choose YIG as the FM layer because of its technological importance, with high $T_c$ ~550 K and extremely low damping coefficient $\alpha$ [26]. When YIG is interfaced with TIs, its good thermal stability minimizes interdiffusion of materials. Through Bi$_2$Se$_3$ thickness dependence study, we observed strong modulation of FMR properties, attributed to the TSS of Bi$_2$Se$_3$. Temperature-dependent study unraveled an effective field parallel to the magnetization direction existing in Bi$_2$Se$_3$/YIG. Such effective field built up as temperature decreased, which was utilized to demonstrate the zero-applied-field FMR of YIG. Furthermore, we identified the signature of MPE of $T_c$ as high as 180 K manifested as enhanced spin pumping in a fluctuating spin system.

Bi$_2$Se$_3$ thin films were grown by molecular beam epitaxy (MBE) [27] on magnetron-sputtered YIG films. The high quality of Bi$_2$Se$_3$/YIG samples in this work is verified by high resolution X-ray diffraction (XRD) and transmission electron microscope (TEM) [28]. To investigate the magnetic properties of Bi$_2$Se$_3$/YIG, room-temperature angle- and frequency-dependent FMR measurements were performed independently using a cavity and co-planar waveguide (CPW), respectively (Figure 1(a) and (b)). For the temperature-dependent FMR, the CPW was mounted in a cryogenic probe station (Lake Shore, CPX-HF), which enable samples to be cooled as low as 5 K. The external field is modulated for lock-in detection in all of the measurements.

The FMR spectra in Figure 1(c) are compared for single layer YIG(12) and Bi$_2$Se$_3$(25)/YIG(12) bilayer (digits denote thickness in nanometer), showing a large shift of resonance field ($H_{res}$) ~317 Oe after the Bi$_2$Se$_3$ growth plus a markedly broadened peak-to-peak width $\Delta H$ for Bi$_2$Se$_3$/YIG. Figure 1(d) shows $H_{res}$ vs applied field angle with respect to surface normal $\theta_H$ for YIG(12) and Bi$_2$Se$_3$(25)/YIG(12). The larger variation of $H_{res}$ with

$\theta_H$ indicates stronger magnetic anisotropy in the bilayer sample. When the applied field was directed in the film plane, clear negative $H_{res}$ shifts induced by Bi$_2$Se$_3$ were observed at all microwave frequency $f$ in Figure 1e. The data can be fitted in the scheme of magnetic thin films having uniaxial perpendicular magnetic anisotropy [28]. Since the XRD results show that the YIG films did not gain additional strain after growing Bi$_2$Se$_3$, the enhanced anisotropy cannot be attributed to changes of magnetocrstalline anisotropy. Defining eff ective demagnetization field $4\pi M_{eff} = 4\pi M_s - H_{an} - H_{int}$, where $4\pi M_s$, $H_{an}$ and $H_{int}$ are the demagnetization field of bare YIG, the magnetocrystalline anisotropy field of YIG, and the anisotropy field induced by Bi$_2$Se$_3$, we obtain $H_{int}$ = -926 and -1005 Oe from Figure 1(d) and 1(e), respectively. The minus sign indicates the additional anisotropy points in the film plane.

The above observations suggested the presence of IMA in Bi$_2$Se$_3$/YIG. To verify this assumption, we systematically varied the thickness of YIG $d_{YIG}$ while fixing the thickness of Bi$_2$Se$_3$. Figure 2(a) presents the $d_{YIG}$ dependence of $4\pi M_{eff}$ for single and bilayer samples. The $4\pi M_{eff}$ of single layer YIG was independent of $d_{YIG}$ varying from 12 to 30 nm. In sharp contrast, $4\pi M_{eff}$ of Bi$_2$Se$_3$/YIG became significantly larger, especially at thinner YIG, which is a feature of an interfacial effect. The $f$- and $\theta_H$-dependent FMR were performed independently to doubly confirm the trends. The IMA can be further characterized by defining the effective anisotropy constant $K_{eff} = (1/2)4\pi M_{eff}M_s = (1/2)(4\pi M_s - H_{an}) + K_i/d_{YIG}$, with the interfacial anisotropy constant $K_i = M_s H_{an} d_{YIG}/2$. The $K_{eff}d_{YIG}$ vs $d_{YIG}$ data in Figure 2(b) is well fitted by a linear function, indicating that the $d_{YIG}$ dependence presented in Figure 2(a) is suitably described by the current form of $K_{eff}$. The intercept obtained by extrapolating the linear function corresponds to $K_i = -0.075$ erg/cm$^2$.

To further investigate the physical origin of the IMA, we next varied the thickness of Bi$_2$Se$_3$ ($d_{BS}$) to see how $K_i$ evolved with $d_{BS}$. Figure 2(c) shows the $d_{BS}$ dependence of $K_i$. Starting from $d_{BS} = 40$ nm sample, the magnitude of $K_i$ went up as $d_{BS}$ decreased. An extremum of $K_i - 0.12 \pm 0.02$ erg/cm$^{-2}$ was reached at $d_{BS} = 7$ nm. An abrupt upturn of $K_i$ occurred in the region $3\,\text{nm} < d_{BS} < 7\,\text{nm}$. The $K_i$ magnitude dropped drastically and exhibited a sign change in the interval. The $K_i$ value of 0.014 erg/cm$^2$ at $d_{BS} = 3\,\text{nm}$ corresponds to weak interfacial perpendicular anisotropy. Based on previous investigation on surface band structure of ultrathin Bi$_2$Se$_3$ [29], $d_{BS} = 6\,\text{nm}$ was identified as the 2D quantum tunneling limit of Bi$_2$Se$_3$. When $d_{BS} < 6\,\text{nm}$, the hybridization of top and bottom TSS developed a gap of the surface states. Spin-resolved photoemission study later showed that the TSS in this regime exhibited decreased *in-plane* spin polarization. The modulated spin texture may lead to the weaker IMA than that in 3D regime [30,31]. We thus divide Figure 2(c) into two regions and correlate the systematic magnetic properties with the surface state band structure. The sharp change of $K_i$ around $d_{BS} = 6\,\text{nm}$ strongly suggests that the IMA in Bi$_2$Se$_3$/YIG is of topological origins.

The $\Delta H$ broadening in FMR spectra after growing Bi$_2$Se$_3$ on YIG indicates additional damping arising from spin-transfer from YIG to Bi$_2$Se$_3$, which is a signature of spin pumping effect [24]. The spin pumping efficiency of an interface can be evaluated by spin mixing conductance $g_{\uparrow\downarrow}$, using the following relation [24],

$$g_{\uparrow\downarrow} = \frac{4\pi M_s d_{YIG}}{g\mu_B}\left(\alpha_{BS/YIG} - \alpha_{YIG}\right) \qquad (1)$$

, where $g$, $\mu_B$, $\alpha_{BS/YIG}$ and $\alpha_{YIG}$ are the Landé $g$ factor, Bohr magneton, the damping coefficient of Bi$_2$Se$_3$/YIG and YIG, respectively. Figure 2(d) displays the $d_{BS}$ dependence of effective spin mixing conductance of Bi$_2$Se$_3$/YIG. Similar to $K_i$ in Figure 2(c), $g_{\uparrow\downarrow}$ had its maximum at $d_{BS} = 7\,\text{nm}$ with a very large value $\sim 2.2 \times 10^{15}$ cm$^{-2}$, about four times larger than that of our Pt/YIG sample indicated by the red dashed line. The inset shows $\Delta H$ vs $f$ data for Bi$_2$Se$_3$ (7)/YIG(13) and YIG(13) fitted by linear functions. One can clearly see a significant change of slope, from which we determined $\alpha_{BS/YIG} - \alpha_{YIG}$ to be 0.014. The large $g_{\uparrow\downarrow}$ of Bi$_2$Se$_3$/YIG implies an efficient spin pumping to an excellent spin sink of Bi$_2$Se$_3$. We again divide Figure 2(d) into two regions according to the surface state band structure. Upon crossing the 2D limit from $d_{BS} = 7$ nm, $g_{\uparrow\downarrow}$ dropped remarkably to $1.7 \times 10^{14}$ at $d_{BS} = 3$ nm. The trend in Figure 2(d) is distinct from that of normal metal (NM)/FM structures. In NM/FM, the $g_{\uparrow\downarrow}$ increases with increasing NM thickness as a result of vanishing spin backflow in thicker NM [32]. It is worth noting that the conducting bulk of $Bi_2Se_3$ can dissipate the spin-pumping-induced spin accumulation at the interface [12,33]. In this regard, the $d_{BS} = 7$ nm sample has the largest weight of surface state contribution to $g_{\uparrow\downarrow}$. Such unconventional $d_{BS}$ dependence of $g_{\uparrow\downarrow}$ suggests that TSS play a dominant role in the damping enhancement.

Since the effects of TSS are expected to enhance at low temperature, we next performed *temperature*-dependent FMR on $Bi_2Se_3$/YIG. Two bilayer samples $Bi_2Se_3(25)$/YIG(15) and $Bi_2Se_3(16)$/YIG(17), and a single layer YIG(23) were measured for comparison. Figure 3(a) and (b) show the $H_{res}$ vs $f$ data at various $T$ for YIG(23) and $Bi_2Se_3(25)$/YIG(15). The $H_{res}$ of both samples show negative shifts at all $f$ with decreasing $T$. The data of YIG(23) can be reproduced by the Kittel equation with increasing $M_s$ of YIG at low $T$. In sharp contrast, $Bi_2Se_3(25)$/YIG(15) exhibited negative intercepts at $H_{res}$, and the intercepts gained its magnitude when the sample was cooled down. This behavior of non-zero intercept is common for all of our $Bi_2Se_3$/YIG samples. To account for the behavior, a phenomenological effective field $H_{eff}$ is added to the Kittel equation, i.e.

$$f = \frac{\gamma}{2\pi}\sqrt{(H_{res} + H_{eff})(H_{res} + H_{eff} + 4\pi M_{eff})}. \qquad (2)$$

The solid lines in Figure 3(b) generated by the modified Kittel equation fitted the experimental data very well.

Figures 3(c) and (d) presents the $T$ dependence of $H_{res}$ and $\Delta H$ for the YIG(23) and two $Bi_2Se_3$/YIG samples. As we lowered $T$, all of the samples had decreasing $H_{res}$, which was viewed as effects of the concurrent increasing $M_{eff}$ and $H_{eff}$ as seen in Figure 3(a) and (b). On the other hand, $\Delta H$ built up with decreasing $T$. We first examine $\Delta H$ of the YIG(23) single layer. The $\Delta H$ remained relatively unchanged with $T$ decreasing from RT, and dramatically increased below 100 K. The pronounced $T$ dependence of $\Delta H$ or $\alpha$ has been explored in various rare-earth iron garnet and was explained by the low $T$ slow-relaxation process via rare-earth elements or $Fe^{2+}$ impurities [34]. For sputtered YIG films, specifically, the increase of $\Delta H$ was less prominent in thicker YIG, indicating that the dominant impurities located near the YIG surface [35]. Distinct from that of YIG(23), the $\Delta H$ progressively increased for the bilayer samples. We were not able to detect FMR signals with $\Delta H$ beyond 100 Oe due to our instrumental limits. However, one can clearly see that, for $Bi_2Se_3(25)$/YIG(15) and $Bi_2Se_3(16)$/YIG(17), the $\Delta H$ broadened due to increased spin pumping at first. For $Bi_2Se_3(25)$/YIG(15), the $\Delta H$ curve gradually leveled off, and intersected with that of YIG(23) at $T \sim 40$ K. The seemingly "anti-damping" by $Bi_2Se_3$ at low $T$ may be related to the modification of the YIG surface chemistry during the $Bi_2Se_3$ deposition. Additional analyses are needed to verify the scenario, which is, however, beyond the scope of this work.

In both $H_{res}$ and $\Delta H$ curves, bump-like features located at $T = 140$ and 180 K (indicated by the arrows) were revealed for $Bi_2Se_3(25)$/YIG(15) and $Bi_2Se_3(16)$/YIG(17), respectively. The bumps are reminiscent of spin pumping in the case of fluctuating magnets, where an enhancement of spin pumping is expected as the spin sink is closed to its magnetic phase transition point [36-38]. In our system, a possibly newly formed magnetic phase would be the interfacial magnetization driven by the proximity effect, namely, $T_c = 140$ and 180 K for our $Bi_2Se_3(25)$/YIG(15) and $Bi_2Se_3(16)$/YIG(17), respectively. In fact, the $T_c$ values of our samples are in good agreement with the reported ones of MPE in TI/YIG systems [6,21]. We did not detect anomalous Hall effect in our samples, which might be obscured by the bulk conduction of $Bi_2Se_3$ in the magneto-transport measurements.

Using Eq. (2), we further determine the T dependence of $4\pi M_{eff}$ and $H_{eff}$ of YIG(23) and the two bilayers samples, as shown in Figures 3(e) and (f) [28]. The $4\pi M_{eff}$ of YIG(23) went larger monotonically as previously discussed, while the $4\pi M_{eff}$ of bilayer samples increased before reaching a maximum of 4000 Oe at $T$ around 150 K, and then decreased slightly at low $T$. With $4\pi M_{eff}^{BS/YIG} - 4\pi M_{eff}^{YIG} \approx -H_{int}$, such $T$ dependence corresponds to decreasing in-plane IMA below 150 K. Note that 150 K is closed to our assumed $T_c$ from MPE at 140 and 180 K. Calculations of total electronic energy at an TI/FM interface show that perpendicular anisotropy is in favor [39], which, in our case, may effectively weaken the in-plane IMA. The decreasing in-plane IMA can also be viewed as a result of modified spin texture by MPE. The analyses thus provide another clue supporting the existence of MPE in our samples. The $H_{eff}$ of bilayer samples, again, show different $T$ evolution than that of the single layer in Figure 3(f). $H_{eff}$ built up with decreasing $T$ in bilayers while the $H_{eff}$ of the YIG single layer was $T$-independent and closed to zero. The positive $H_{eff}$ corresponds to an effective field parallel to the magnetization vector ***M*** and resembles the exchange bias field. However, we did not observe shifts of magnetization hysteresis loop characteristic of an exchange bias effect [28,40]. The $H_{eff}$ present in FMR measurement suggests it may come from spin imbalance at the interface [41].

    Finally, we demonstrated that the large IMA and $H_{eff}$ in Bi$_2$Se$_3$/YIG are strong enough to induce FMR without an $H_{ext}$, which we term *zero-field FMR*. Figure 4(a) displays $T$ evolution of FMR first derivative spectra of Bi$_2$Se$_3$(25)/YIG(15) at $f = 3.5$ GHz. The spectral shape started to deform when the $H_{res}$ was approaching zero. The sudden twists at $H_{ext}$ ~ +30 (-30) for positive (negative) field sweep arose from magnetization switching of YIG, and therefore led to hysteric spectra. The two spectra merged at 25 K and then separated again when $T$ was further decreased. Figure 4(b) shows the microwave absorption intensity $I$ spectra with positive field sweeps. We traced the position of $I$ spectrum $H_{peak}$ using the red dashed line, and found it coincided with zero $H_{ext}$ at the zero-field FMR temperature $T_0$ ~25 K. Below 25 K, $H_{peak}$ moved across the origin and one needed to reverse $H_{ext}$ to counter the internal effective field comprised of the demagnetization field $4\pi M_s$, $H_{int}$ and $H_{eff}$ (Figure 4(e)).

Note that the presence of $H_{int}$ alone would be inadequate to realize zero-field FMR. Only when $H_{eff}$ is finite would the system exhibit non-zero intercepts as we have seen in Figure 3(b). We further calculate $T_0$ as a function of microwave excitation frequency $f$ (Figure 4(f)) using Eq. (2) and the extracted $H_{eff}$ of Figure 3(f). We obtain that, with finite $H_{eff}$ persisting up to room temperature, zero-field FMR can be realized at high $T$ provided $f$ is sufficiently low. However, we emphasize that it's advantageous for YIG to be microwave-excited above 3 GHz. When $f < 3$ GHz, parasitic effects such as three-magnon splitting [42,43] take place and significantly decrease the microwave absorption in YIG. Here, we demonstrate that the strong exchange coupling between $Bi_2Se_3$ and YIG gave rise to zero-field FMR in the feasible high frequency operation regime of YIG. Further improvement of interface quality of $Bi_2Se_3$/YIG is expected to raise $H_{eff}$ and $T_0$ for room temperature, field free spintronic application.

In summary, we investigate the magnetization dynamics of YIG in the presence of interfacial exchange coupling and TSS of $Bi_2Se_3$. The significantly modulated magnetization dynamics at room temperature is shown to be TSS-originated through the $Bi_2Se_3$ thickness dependence study. It can be expected that the strong coupling between $Bi_2Se_3$ and YIG will modify interface band structures and even the spin texture of TSS. The underlying mechanism of the coexisting large in-plane IMA and pronouncedly enhanced damping calls for further theoretical modeling and understanding. The temperature-dependent FMR results revealed rich information, including signatures of MPE and increasing exchange effective field at low temperature. Our study is an important step toward realization of topological spintronics.

We would like to thank helpful discussion with Profs. Mingzhong Wu and Hsin Lin. The work is supported by MOST 105-2112-M-007-014-MY3, 105-2112-M-002-022-MY3, 106-2112-M-002-010-MY3, 102-2112-M-002-022-MY3, and 105-2112-M-001-031-MY3 of the Ministry of Science and Technology in Taiwan.

Figure 1.

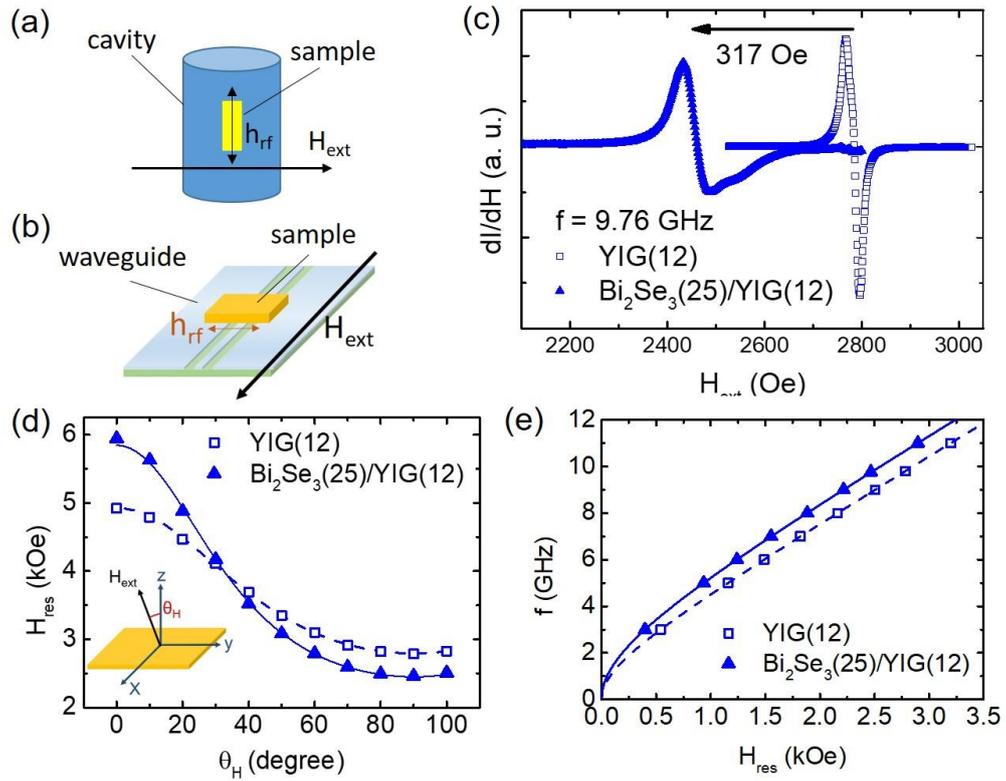

FIG. 1. (a), (b) FMR using the cavity and co-planar configuration for angle- and frequency-dependent study, respectively. A dc external field $H_{ext}$ was applied and $h_{rf}$ denotes the microwave field. (c) FMR spectra of $Bi_2Se_3(25)/YIG(12)$ and YIG(12) measured by the cavity. (d), (e) $\theta_H$ and $f$ dependence of $H_{res}$ of $Bi_2Se_3(25)/YIG(12)$ and YIG(12), respectively.

Figure 2.

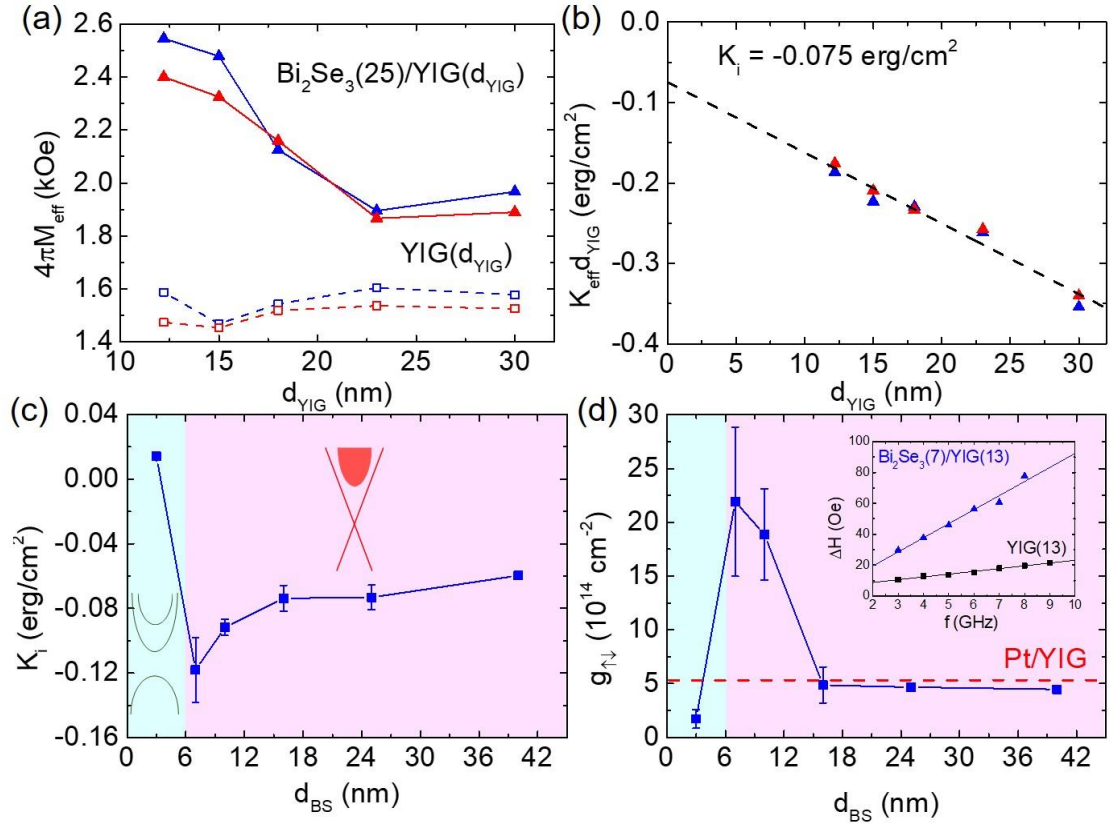

FIG. 2. (a) The $d_{YIG}$ dependence of $4\pi M_{eff}$ of Bi$_2$Se$_3$/YIG (solid triangles) and YIG (hollow squares) obtained from $\theta_H$ (red) and $f$ (blue) dependent FMR. (b) The $K_{eff}d_{YIG}$ vs $d_{YIG}$ plot for determining $K_i$ using a linear fit. The intercept of the y-axis corresponds to the $K_i$ value. (c) $d_{BS}$ dependence of $K_i$. The figure is divided into two regions. For $d_{BS} > 6$ nm, the Dirac cone of TSS is intact, with the Fermi level located in the bulk conduction band. For $d_{BS} < 6$ nm, a gap and quantum well states form. (d) The $d_{BS}$ dependence of spin mixing conductance $g_{\uparrow\downarrow}$. The inset shows $\Delta H$ as a function of Bi$_2$Se$_3$(7)/YIG(13) and YIG(13) for calculating $g_{\uparrow\downarrow}$. The red dashed line indicates the typical value of $g_{\uparrow\downarrow}$ of Pt/YIG.

Figure 3.

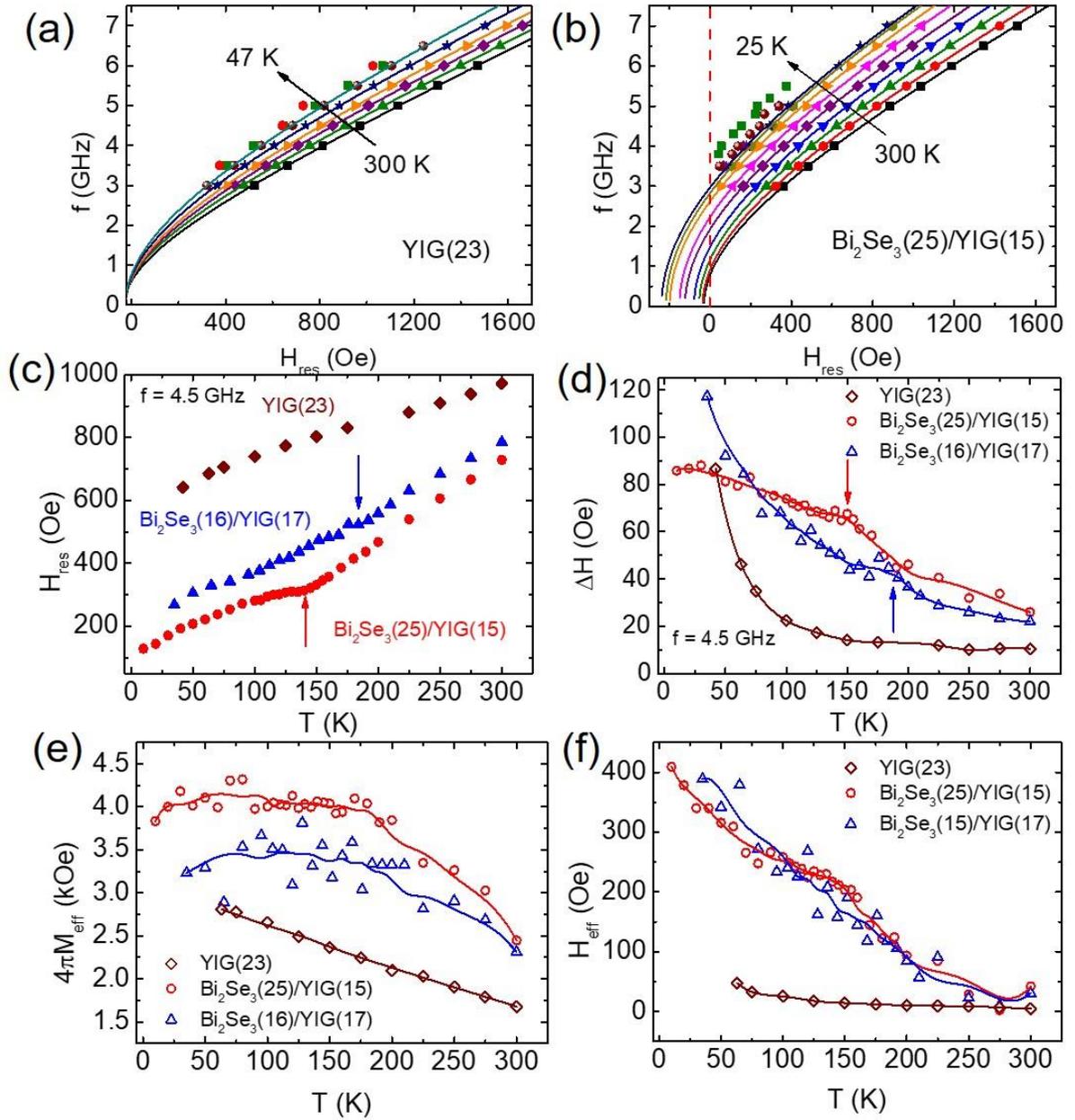

FIG. 3. (a), (b) $f$ vs $H_{res}$ data for various $T$ for YIG(23) and Bi$_2$Se$_3$(25)/YIG(15), respectively. Solids lines are fitted curves using Eq. (2). (c), (d), (e), and (f) $T$ dependence of $H_{res}$, $\Delta H$, $4\pi M_{eff}$ and $H_{eff}$ of the one YIG (23) single layer and two Bi$_2$Se$_3$/YIG bilayer samples, Bi$_2$Se$_3$(25)/YIG(15) and Bi$_2$Se$_3$(16)/YIG(17), respectively. The arrows in (c) and (d) denote the position of the "bumps". Solids line are guide of eyes obtained by properly smoothing the experimental data.

Figure 4.

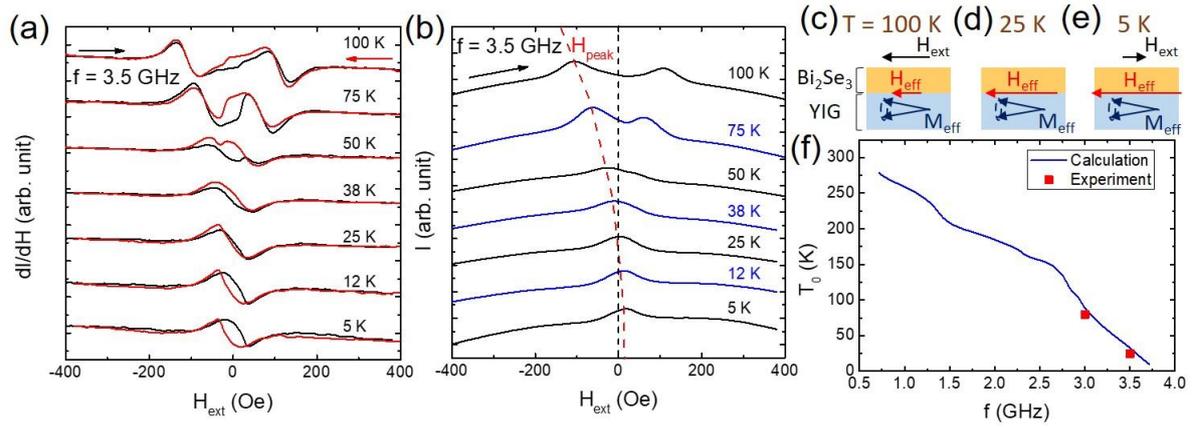

FIG. 4. (a), (b) FMR first derivative and microwave absorption spectra for various $T$, respectively. The arrows indicate the $H_{ext}$ sweep direction. The dashed line in (b) traces the $T$ evolution of the absorption peak $H_{peak}$. (c), (d), and (e) Schematics of the $Bi_2Se_3$/YIG sample when $T = 100$, 25, and 5 K, respectively. (f) Zero-field FMR temperature $T_0$ as a function of $f$.

# Supplemental materials

**Growth and FMR characteristics of YIG films**

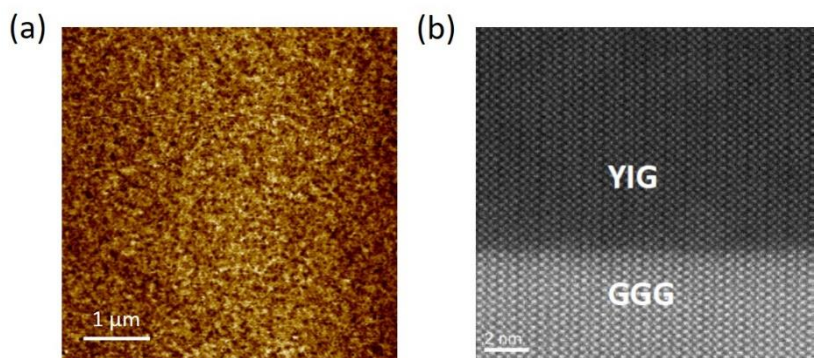

FIG. S1. (a) and (b) AFM surface image and HAADF-STEM image of YIG/GGG.

The YIG thin films were deposited on (111)-oriented GGG substrates by off-axis sputtering at room temperature. The GGG(111) substrates were first ultrasonically cleaned in order of acetone, ethanol and DI-water before mounted in a sputtering chamber with the base pressure of $2 \times 10^{-7}$ mTorr. For YIG deposition, a 2-inch YIG target was sputtered with the following conditions: an applied rf power of 75 W, an Ar pressure of 50 mtorr and a growth rate of 0.6 nm/min. The samples were then annealed at 800°C with $O_2$ pressure of 11.5 mtorr for 3 hours. Fig. S1(a) displays the atomic force microscopy (AFM) image of the YIG surface, showing a flat surface with roughness of 0.19 nm. Fig. S1(b) shows the high-angle annular dark-field (HAADF) image of YIG/GGG. The YIG thin film was epitaxially grown on the GGG substrate with excellent crystallinity. No crystal defects were observed at the YIG bulk and YIG/GGG interface.

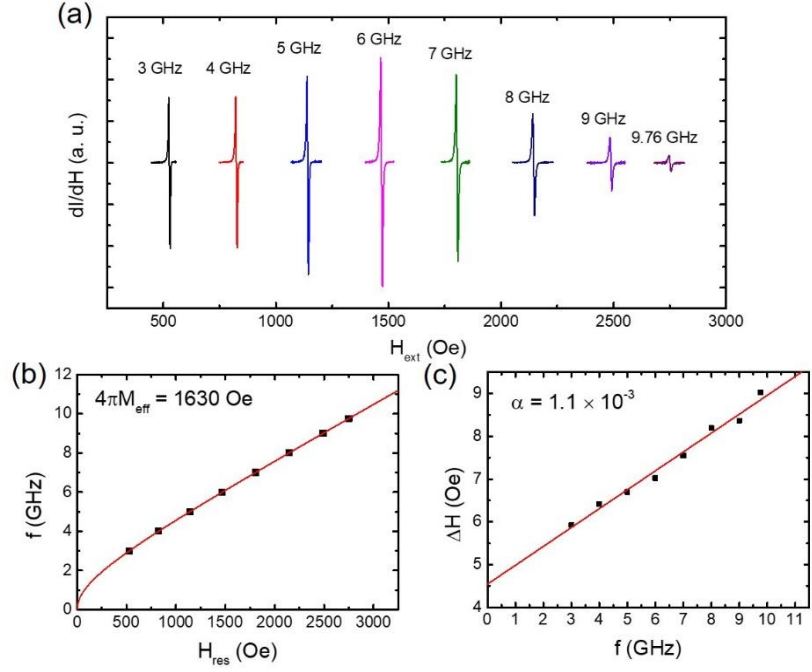

FIG. S2. (a) FMR first-derivative spectra of 23 nm YIG at various frequencies. (b) $f$ vs $H_{res}$ data fitted to the Kittel equation (red line). (c) $\Delta H$ as a function of $f$ data. From the linear fit (red line) the $\alpha$ value of the sample is obtained.

Fig. S2(a) shows the representative FMR data of our YIG film measured by coplanar waveguide. The FMR spectra exhibit Lorentzian lineshape at all measured frequencies ranging from 3 to 9.76 GHz. To determine the $4\pi M_{eff}$ and $\alpha$, the resonance fields $H_{res}$ and peak-to-peak widths $\Delta H$ of these spectra were plotted as a function of $f$ as shown in Fig. S2(b) and (c), respectively. We obtain the $4\pi M_{eff}$ of our 23 nm YIG to be 1630 Oe. We note that the $4\pi M_{eff}$ value is lower than the reported values of the YIG prepared by either sputtering or pulsed laser deposition [S1-3]. The difference may come from growth conditions dependent on different systems. For determination of $\alpha$, the data in Fig. S2(c) is fitted to the following equation,

$$\Delta H = \Delta H_0 + \frac{4\pi f \alpha}{\sqrt{3}\gamma}. \qquad (S1)$$

Here, $\Delta H_0$ and $\gamma$ are the inhomogeneous broadening and gyromagnetic ratio. The linear fit in Fig. S2(c) corresponds to $\alpha = 1.1 \times 10^{-3}$.

## Structural characterizations of Bi$_2$Se$_3$/YIG heterostructures

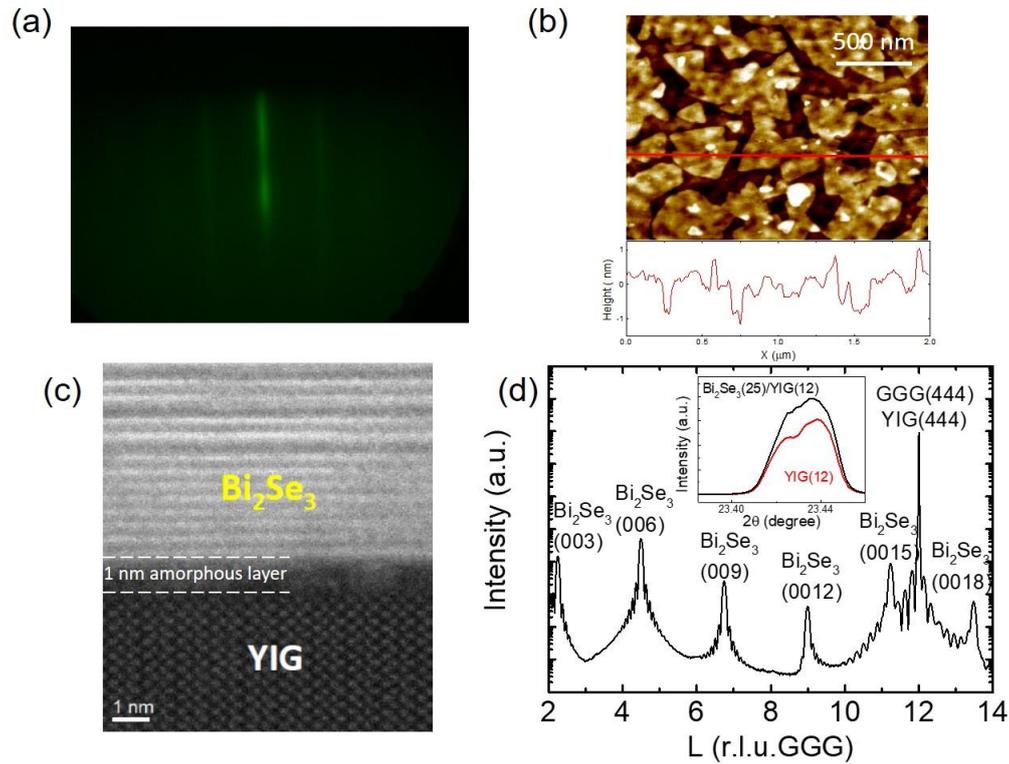

FIG. S3. (a) RHEED patterns of MBE grown 7 QL Bi$_2$Se$_3$ on YIG/GGG(111) substrates. (b) AFM image of a 7 QL Bi$_2$Se$_3$. (c) HAADF-STEM image of Bi$_2$Se$_3$/YIG/GGG heterostructures. (d) SR-XRD of our Bi$_2$Se$_3$(25)/YIG(12) sample. Clear Pendellösung fringes of YIG and Bi$_2$Se$_3$ indicates excellent crystallinity. The inset shows the radial scans of YIG before and after Bi$_2$Se$_3$ growth.

The YIG/GGG samples were annealed at 450°C in the MBE growth chamber for 30 min prior to Bi$_2$Se$_3$ growth at 280°C [27]. The base pressure of the system was kept about $2 \times 10^{-10}$ Torr. Elemental Bi (7N) and Se (7N) were evaporated from regular effusion cells. As shown in Fig. S3a, streaky reflection high-energy electron diffraction (RHEED) patterns of Bi$_2$Se$_3$ were observed. Fig. S3b displays the surface morphology of 7 QL Bi$_2$Se$_3$ taken by atomic force microscopy (AFM). The image shows layer-by-layer growth of Bi$_2$Se$_3$ with the step heights ~ 1nm, which corresponds to the thickness of 1 QL. The layer structure of Bi$_2$Se$_3$ was also revealed by the HAADF image shown in Fig. S3c. Despite the high quality growth of Bi$_2$Se$_3$, an amorphous interfacial layer of ~ 1 nm formed. The excellent crystallinity of our samples was verified by clear

Pendellösung fringes of the synchrotron radiation x-ray diffraction (SR-XRD) data shown in Fig. S3(d). In particular, the radial scans data of YIG/GGG(22-4) before and after growing Bi₂Se₃ are perfectly matched, indicating the absence of Bi₂Se₃-induced strains in YIG that might contribute additional magnetic anisotropy [S4].

**Analyses of magnetic anisotropy**

We express the free energy density E of the system as

$$E = -\mathbf{M} \cdot \mathbf{H} + \frac{1}{2}M_s(4\pi M_s - H_{an} - H_{int})\cos^2\theta_M \qquad (S2)$$

, where $\mathbf{M}$, $\mathbf{H}$, $M_s$, $H_{an}$ and $\theta_M$ are magnetization vector, applied field vector, saturation magnetization, the anisotropy field induced by Bi₂Se₃ and magnetization angle with respect to the surface normal, respectively. We further define the effective demagnetization field $4\pi M_{eff} = 4\pi M_s - H_{an} - H_{int}$, where $H_{an}$ is the magnetocrystalline anisotropy field of sputtered YIG and $H_{int}$ is the interfacial anisotropy field. We have $H_{int} = 0$ for the YIG single layer by definition. The first term of Eq. (S2) is the Zeeman energy and the second term accounts for uniaxial out-of-plane anisotropy. Here, we neglect higher order terms that are relatively small for a strain-free cubic system. The $H_{res}$ can be calculated by minimizing $E$ and, at the equilibrium angle of $\mathbf{M}$, solving the Smit-Beljers equation [S5],

$$\left(\frac{\omega}{\gamma}\right)^2 = \frac{1}{M^2 \sin^2\theta_M}\left[\frac{\partial^2 E}{\partial \theta_M^2}\frac{\partial^2 E}{\partial \varphi_M^2} - \left(\frac{\partial^2 E}{\partial \theta_M \partial \varphi_M}\right)^2\right]. \qquad (S3)$$

As $\theta_H = \pi/2$, the FMR conditions reduce to the Kittel equation $f = \frac{\gamma}{2\pi}\sqrt{H_{res}(H_{res} + 4\pi M_{eff})}$.

We can safely assume that the $H_{an}$ did not change before and after the growth of Bi₂Se₃ based on Figure S3(d). With this in mind, we further notice that $K_i$ can be alternatively expressed as $(1/2)(4\pi M_{eff}^{YIG} - 4\pi M_{eff}^{BS/YIG})M_s d_{YIG}$, where

$4\pi M_{eff}^{YIG}$ ($4\pi M_{eff}^{BS/YIG}$) represents the $4\pi M_{eff}$ of YIG (Bi$_2$Se$_3$/YIG) for a specific $d_{YIG}$. The calculated $K_i$ using this expression for $d_{BS}$ = 25 nm gives an average of -0.068 erg/cm$^2$, in good agreement with a $K_i$ of -0.075 erg/cm$^2$ obtained from the linear fit in Figure 2(b).

**Temperature-dependent FMR spectra of YIG and Bi$_2$Se$_3$/YIG**

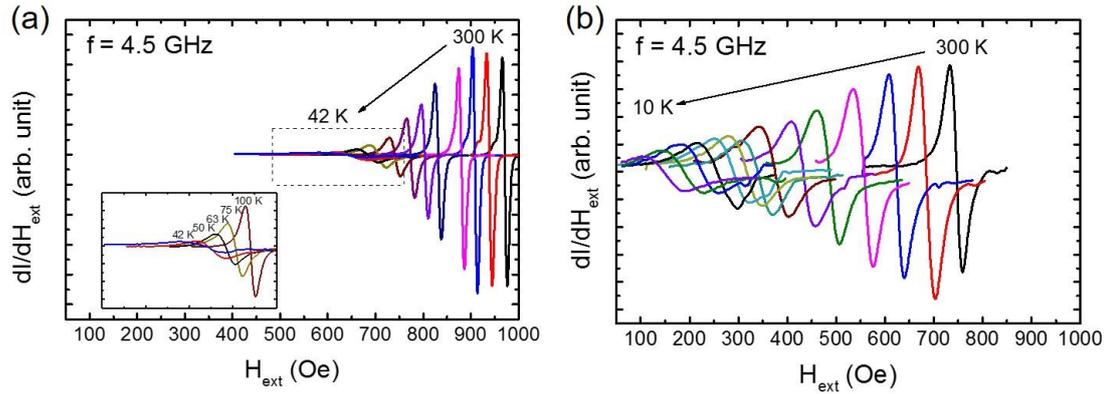

FIG. S4. Temperature-dependent FMR first-derivative spectra of (a) YIG(23) and (b) Bi$_2$Se$_3$(25)/YIG(15). The $\Delta H$ increased with decreasing $T$, accompanied by decreased $dI/dH_{ext}$ peak magnitudes due to enhanced damping. The $H_{ext}$ scale are fixed to clearly show the pronounced changes of $H_{res}$ and $\Delta H$ induced by Bi$_2$Se$_3$.

**Extraction of $4\pi M_{eff}$ and $H_{eff}$ from temperature dependence of $H_{res}$**

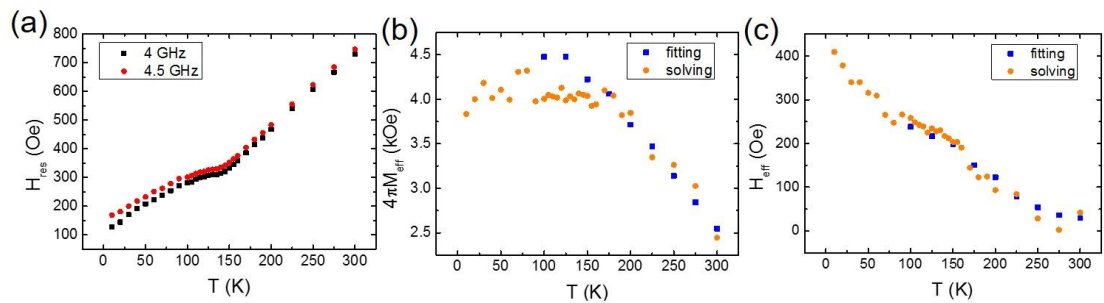

FIG. S5. (a) $H_{res}$ vs $T$ data of Bi$_2$Se$_3$(25)/YIG(15) measured at 4 and 4.5 GHz. (b) and (c) Comparison of extracted $4\pi M_{eff}$ and $H_{eff}$ by the fitting and solving method.

Since the effective damping constant of YIG and Bi$_2$Se$_3$/YIG increased pronouncedly at low $T$, the weakened FMR signal was inevitably accompanied by larger uncertainties of measured $H_{res}$. The uncertainties are even more serious when

$T < 150$ K and $f > 5$ GHz for our instruments. Fitting the data including points in the $f > 5$ GHz region to the Kittel equation gives large errors of $4\pi M_{eff}$ and $H_{eff}$, which obscure the temperature dependency of these two quantities. To reduce the uncertainties in the process of extracting $4\pi M_{eff}$ and $H_{eff}$, we focused on the FMR data for $f = 4$ and 4.5 GHz, from which we obtained $H_{res}$ with satisfactory accuracy (Fig. S5(a)). The Kittel equation can be arranged in the following form,

$$H_{eff}^2 + (2H_{res} + 4\pi M_{eff})H_{eff} + \left(4\pi M_{eff}H_{res} + H_{res}^2 - \frac{4\pi^2 f^2}{\gamma^2}\right) = 0. \quad (S3)$$

With $\gamma = 1.77 \times 10^{11}$ t$^{-1}$s$^{-1}$, the two sets of data in Fig. S5a provided us with sufficient information to explicitly solve the second-order equation. Fig. S5b and S5c compare the results of "fitting" and "solving" the Kittel equation. For $T > 150$ K, where FMR can be accurately measured up to 7 GHz, the $4\pi M_{eff}$ and $H_{eff}$ obtained from solving and fitting method agree well, demonstrating the reliability of the solving method.

**Temperature dependence of magnetization hysteresis loop**

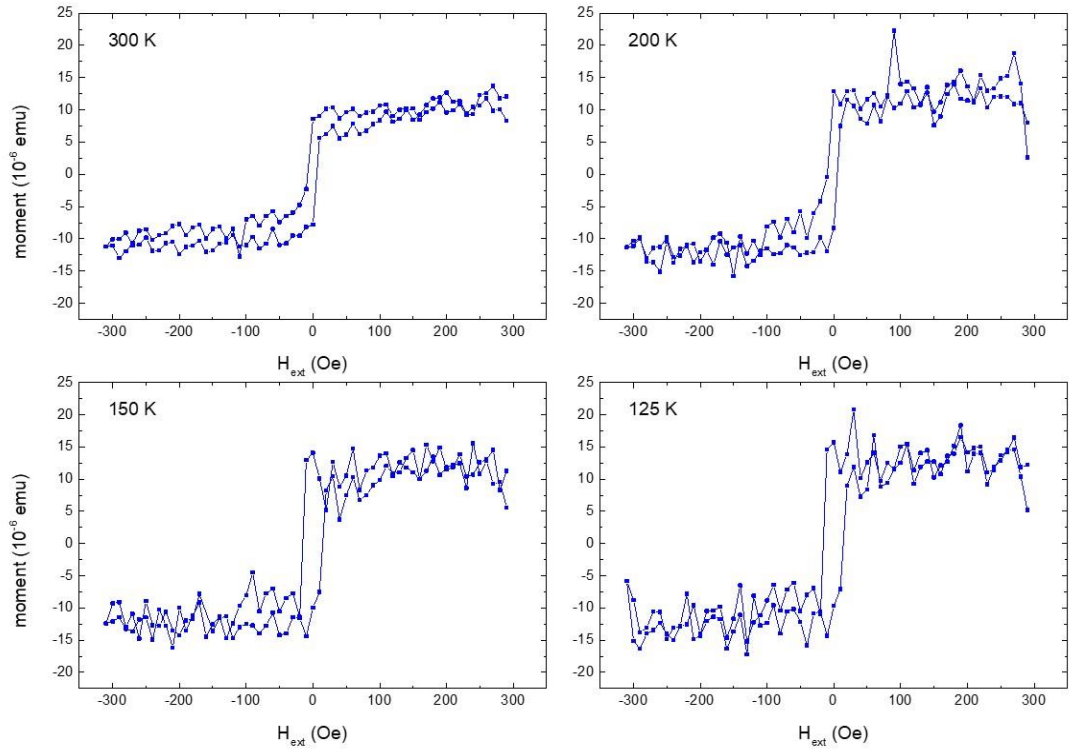

FIG. S6. Temperature dependence of magnetization hysteresis loop of $Bi_2Se_3(16)/YIG(17)$ measured by a SQUID magnetometer. The paramagnetic background of the GGG substrate has been subtracted. No shifts of hysteresis loops were observed.